\documentclass[prl,aps,twocolumn,showpacs,preprintnumbers,superscriptaddress]{revtex4}
\usepackage{graphicx}% Include figure files
\usepackage{dcolumn}% Align table columns on decimal point
\usepackage{bm}% bold math
\usepackage{amssymb}
\usepackage{amsmath}
\usepackage{times}
\usepackage[nooneline]{subfigure}

\newcommand\eps{\epsilon}

\begin{document}

%\title{Additive noise induced state transitions and universality in
%the noisy Kuramoto-Sivashinksy equation}
\title{Noise induced state transitions, intermittency and universality in
the noisy Kuramoto-Sivashinsky equation}
\author{M. Pradas}%\email{}
\affiliation{Department of Chemical Engineering, Imperial
College London, London, SW7 2AZ, UK}

\author{D. Tseluiko}%\email{}
\affiliation{School of Mathematics, Loughborough University, Leicestershire,
LE11 3TU, UK}

\author{S. Kalliadasis}%\email{}
\affiliation{Department of Chemical Engineering, Imperial
College London, London, SW7 2AZ, UK}

\author{D. T. Papageorgiou}%\email{}
\affiliation{Department of Mathematics, Imperial
College London, London, SW7 2AZ, UK}

\author{G. A. Pavliotis}%\email{}
\affiliation{Department of Mathematics, Imperial
College London, London, SW7 2AZ, UK}

\date{\today}

\begin{abstract}
Consider the effect of pure additive noise on the long-time
dynamics of the noisy Kuramoto-Sivashinsky (KS) equation
close to the instability onset. 
When the noise acts only on the first stable mode 
(highly degenerate),
%in the sense that it acts only on the first stable mode, 
the KS solution undergoes several state transitions, %{\bf{between differentstates}}, 
including critical on-off 
intermittency %that is eventually stabilized
and stabalized states, 
as the noise strength increases.
Similar results are obtained with the
%{\bf{noisy}} 
Burgers equation. Such noise-induced transitions
are completely characterized through critical exponents, obtaining 
the same universality class for both %{\bf{the KS and the Burgers}}
equations, and rigorously explained  using multiscale techniques.
\end{abstract}

 \maketitle

Most physical and technological settings are subject to random
fluctuations, which
are responsible for many intriguing and surprising phenomena~\cite{Horsthemke_Lefever_1984}.
These settings are often described by model
spatially extended systems (SES), i.e.~infinite-dimensional
dynamical systems with space-time dependence and 
some stochastic forcing~\cite{Sagues.etal_2007}.
A widely studied example
is the transition between different observed system
\emph{states} as the noise strength is increased beyond a critical
value. For both pure temporal dynamical systems and fully nonlinear
SES, it is well known that noise-induced transitions are due to
multiplicative noise, i.e.~noise whose amplitude depends on the
fluctuating
variable~\cite{Horsthemke_Lefever_1984,Mackey.etal_1990}. The
presence of an additive noise, i.e.~noise that does not depend on the
state of the system, in addition to multiplicative one, has been
shown to induce other phase transitions~\cite{Landa.etal_1998} while
recently it has been shown that \emph{pure additive noise},
i.e.~thermal fluctuations, can stabilize linearly unstable solutions of
SES~\cite{Blomker.etal_2007, Hutt.etal_2007}. However, a
satisfactory and systematic description of the effects of thermal
fluctuations on SES as well as a quantitative description of such
effects in terms of critical state transitions is still lacking.

In this Letter we report analytical and numerical evidence of pure
additive noise-induced transitions in SES. As a main case study, we
consider the noisy KS equation close to the primary bifurcation. We
observe numerically a number of critical transitions by increasing
the noise strength, including on-off intermittency, a crucial
universal feature of many nonlinear systems close to criticality
reflecting a transition from order/coherence to a disordered state
(hence understanding the statistical properties of intermittency is
crucial for the characterization of this transition). Our numerical
observations can be fully explained in the context of a multiscale
theory for SES.

\emph{Noise in weakly nonlinear evolution equations.-}
We consider the noisy KS equation
\begin{equation}\label{Eq:KS}
\partial_t u=-(\partial_x^2+\nu\partial_x^4)u-
u\partial_x u+\tilde{\sigma}\xi,
\end{equation}
normalized to 
$2\pi$-domains so that $0<\nu=(\pi/L)^2$, where $2L$ is the original length
of the system, and with either homogeneous Dirichlet Boundary
Conditions (DBC) or Periodic Boundary Bonditions (PBC).
Equation~\eqref{Eq:KS}, with and without the noise term, has
attracted a lot of attention since it appears in a wide variety of
physical phenomena and applications and it also serves as a
canonical reference system of SES exhibiting spatiotemporal chaos
or dissipative turbulence, e.g. reaction-diffusion systems and
interfacial instabilities in fluid flows~\cite{KSapplications}.

\begin{figure}
\centering
\includegraphics[width=0.5\textwidth]{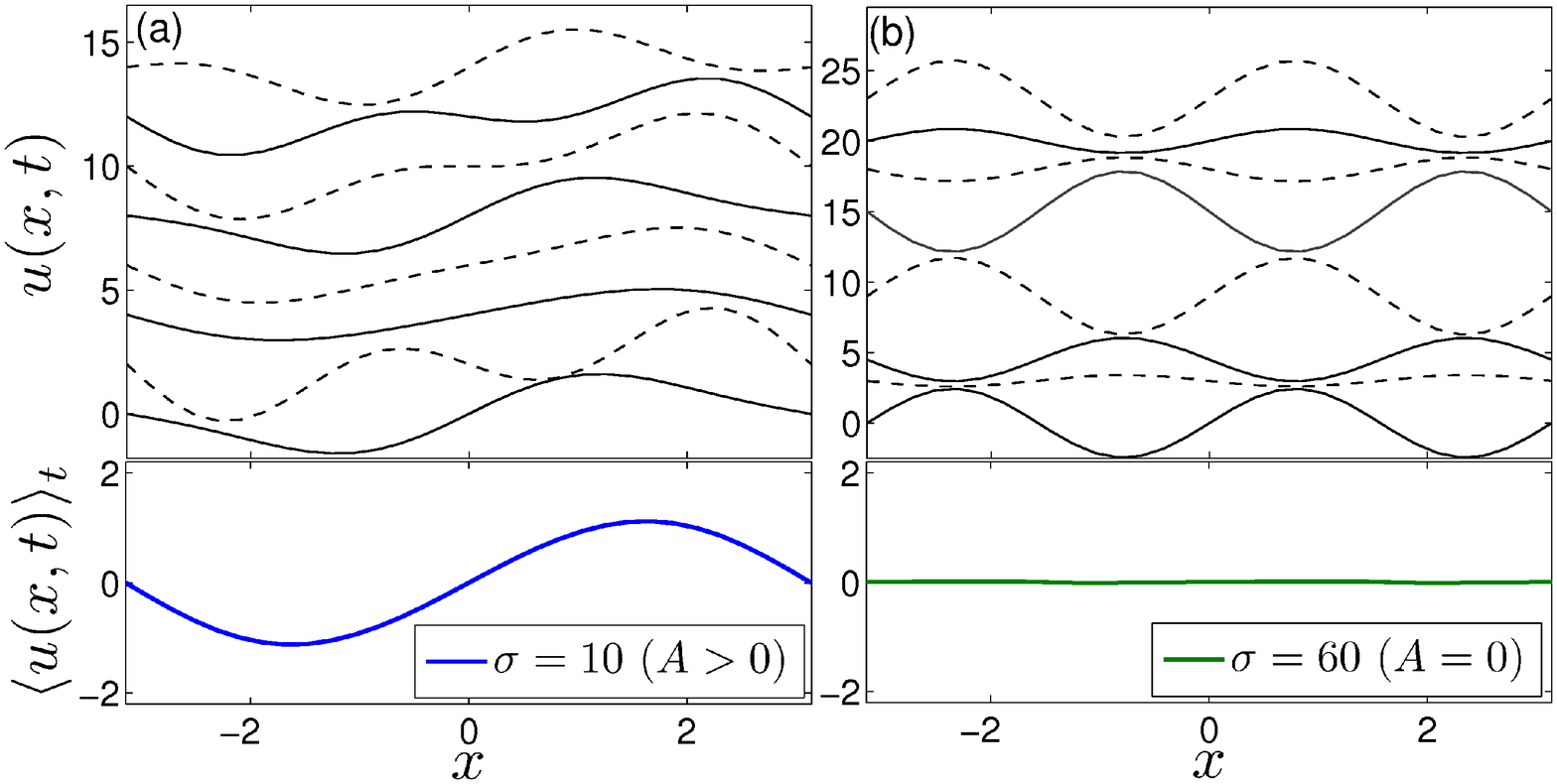}
\caption{(Color online) Top panels show typical spatio-temporal
evolution of the noisy KS equation solved with DBC for
$\epsilon=0.1$ and $\sigma=10$ (a) and $60$ (b) at time intervals
$\Delta t=100$ depicted as solid-dashed lines. For clarity, the
curves are arbitrarily shifted in the vertical direction. Bottom
panels show the corresponding time average of
$u(x,t)$.}\label{Fig:fig1}
\end{figure}

We shall assume throughout zero-mean solutions, 
and we study  a randomly perturbed regime close to criticality by
slightly increasing the domain size as $L=\pi\sqrt{1+\epsilon^2}$ so
that we write $\nu=1-\epsilon^2$ and
$\tilde{\sigma}=\epsilon\sigma$,  where $\sigma$ represents the
strength of the noise, with $\epsilon$ being a bifurcation
parameter; if $\epsilon=0$ all modes, except the neutral one, are stable with the system
approaching its rest state as $t \to \infty$, and for
$0<\epsilon<1$, a bifurcation occurs leading to a finite number of
linearly unstable modes (in fact $\lfloor 1/\sqrt{1-\epsilon^2}\rfloor$
of them). The field $u$ can then  be projected onto the set of
eigenfunctions $\left\{e_k(x)\right\}$ for $k=1, 2, \ldots$ of the
linear operator $\mathcal{L}=-\partial_x^2-\partial_x^4$, such that
$u(x,t)=\sum_k\hat{u}_k(t)e_k(x)$, and in the limit $\epsilon
\rightarrow 0$, only one single mode, namely $\hat{u}_1(t)$, will be
unstable. We are interested in the dynamics of $\hat{u}_1(t)$ when
the stable modes, $\hat{u}_k(t)$ for $k\geq 2$, are randomly forced,
and in particular we focus on the case when only the first stable
mode ($\hat{u}_2$) is perturbed, so that the noise term in
Eq.~(\ref{Eq:KS}) is written as $\xi(x,t)=\hat{\xi}_2(t)e_2(x)$,
where $\hat{\xi}_2(t)$ is some uncorrelated Gaussian noise. This
``highly degenerate noise" may give rise to a stabilization process
of the unstable mode $\hat{u}_1$~\cite{Blomker.etal_2007}.
Typical snapshots of the spatio-temporal evolution of
Eq.~(\ref{Eq:KS}) subject to DBC with
$\sigma=10$ and $60$ %$\epsilon=0.025$
are depicted in Fig.~\ref{Fig:fig1}.
The dynamic evolution of the first mode amplitude, $A(t)\equiv \vert
\hat{u}_1(t)\vert$ is calculated for different noise strengths and
boundary conditions.

\begin{figure}
\centering
\includegraphics[width=0.5\textwidth]{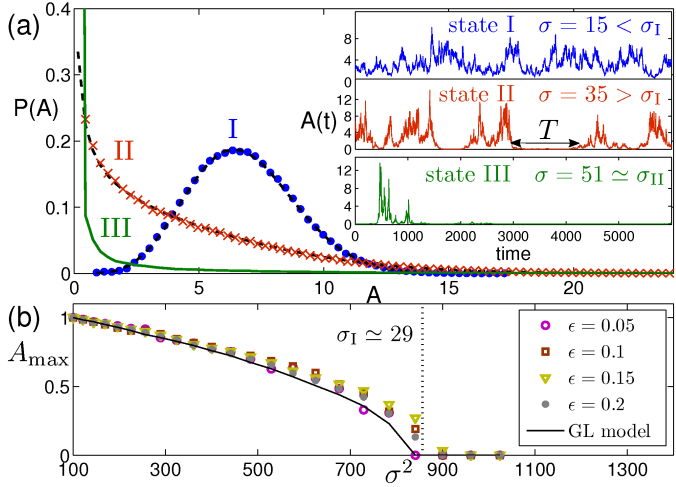}
\caption{(Color online) Numerical results for Eq.~(\ref{Eq:KS})
integrated on a $[-\pi,\pi]$ domain with DBC. (a) PDF of the
first-mode amplitude $A(t)=\vert \hat{u}_1(t)\vert$ for $\sigma=15$
($\bullet$), $35$ ($\times$), and $51$ (green solid line), with
$\epsilon=0.1$. Dashed lines correspond to a data fit using $P(A)=N
A^{\alpha_1}\exp{(-\delta A^2)}$, where the fitted value $\alpha_1$
is related to Eq.~(\ref{Eq:paramters}). The inset depicts typical fluctuations of
the amplitude at each of the three states discussed in the text. (b)
Maximum of the PDF as a function of $\sigma^2$ for different values
of $\epsilon$. The solid line corresponds to the theoretical
solution  obtained by solving numerically Eq.~(\ref{Eq:GL}) with the
coefficients of Eq.~(\ref{Eq:coeff KS}). All curves have been
normalized to the corresponding minimum value at
$\sigma=10$.}\label{Fig:fig2}
\end{figure}

In the case of DBC, the Probability Density Function (PDF) of $A(t)$
calculated for different values of  $\sigma$ is shown in
Fig.~\ref{Fig:fig2}$a$. For $\sigma =15$ the amplitude is
characterized by finite fluctuations that never reach zero (state~I,
top panel in the inset of Fig.~\ref{Fig:fig2}$a$), and the PDF can
be fitted to a function of the form $P(A)=N
A^{\alpha_1}\exp{(-\delta A^2)}$. For $\sigma = 35$, the noise is
strong enough to alter the behavior of the PDF, shifting its maximum
position, $A_\mathrm{max}$, which can now reach zero (state~II). We 
characterize this transition between I and II by
computing $A_\mathrm{max}$ for different  $\sigma$ 
(cf.~Fig.~\ref{Fig:fig2}$b$), obtaining 
$A_\mathrm{max}\sim \vert\sigma^2-850\vert^{1/2}$ which gives a
critical value of $\sigma_\mathrm{I}\simeq 29$. Finally, as we
increase $\sigma$ up to the value  $\sigma=51$ the
first mode is completely stabilized (state~III), and the
fluctuations eventually reach zero, defining a second critical
transition at $\sigma_\mathrm{II}\simeq 50$. Such
stabilization process can be clearly observed when the solution
$u(x,t)$ is averaged over time (cf.~Fig.\ref{Fig:fig1}).

\begin{figure}
\centering
\includegraphics[width=0.5\textwidth]{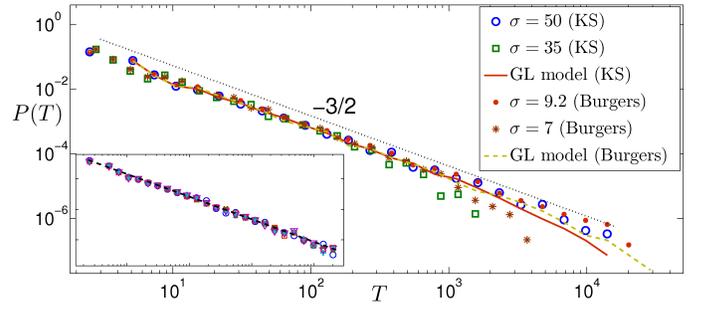}
\caption{(Color online) PDF of the waiting times $T$ between two
consecutive bursts observed in the on-off intermittent state II
corresponding to the DBC case. We solved both the KS equation, with
$\sigma=35$ and $50$, and the Burgers equation, with $\sigma=9.2$
and $\sigma=7$, by using $\epsilon=0.1$. The solid and dashed lines
correspond to the numerical solutions of the GL model,
Eq.~(\ref{Eq:GL}), by using the KS coefficients of
Eq.~(\ref{Eq:coeff KS}) with $\sigma=50$, and the corresponding
Burgers coefficients with $\sigma=9$, respectively. The dotted line
is a data fit to $P(T)\sim T^{-\tau}$ with $\tau=1.50\pm 0.01$.  The
inset shows the waiting times PDF in the KS equation with $\sigma
=50$ and different values for the threshold, namely
$c_\mathrm{th}=0.05$, $0.1$, $0.2$, $0.4$, and
$0.8$.}\label{Fig:fig3}
\end{figure}

The middle panel in the inset of Fig.~\ref{Fig:fig2}$a$ demonstrates
that state~II is characterized by an \emph{on-off intermittent}
behavior of the amplitude fluctuations. Such intermittency can
be characterized by studying the PDF of the waiting times $T$
between two consecutive bursts, defined as large fluctuations above
a given threshold, i.e.~$A(t)>
c_\mathrm{th}$~\cite{Heagy.etal_1994}. Figure \ref{Fig:fig3} shows
the numerical results obtained by using two noise
strengths. For $\sigma=50$, %which is
close to the second critical point ($\sigma_\mathrm{II}$), the PDF of %the waiting times
$T$ is given by %a pure power-law,
$P(T)\sim T^{-\tau}$ with $\tau=3/2$. Interestingly, this exponent
has been ubiquitously found in many other physical systems that
display avalanche or intermittent dynamics close to criticality,
including neuronal activity in cortex, electroconvection of nematic
liquid crystals or fluid flow in porous media~\cite{exponent_3_2}.
For $\sigma=35$, far from $\sigma_\mathrm{II}$, the
PDF is exponentially corrected.
These results %around the critical point ($\sigma\simeq\sigma_\mathrm{II}$)
do not depend on the choice of the threshold value $c_\mathrm{th}$ (see inset of
Fig.~\ref{Fig:fig3}).
\begin{figure}
\centering
\includegraphics[width=0.5\textwidth]{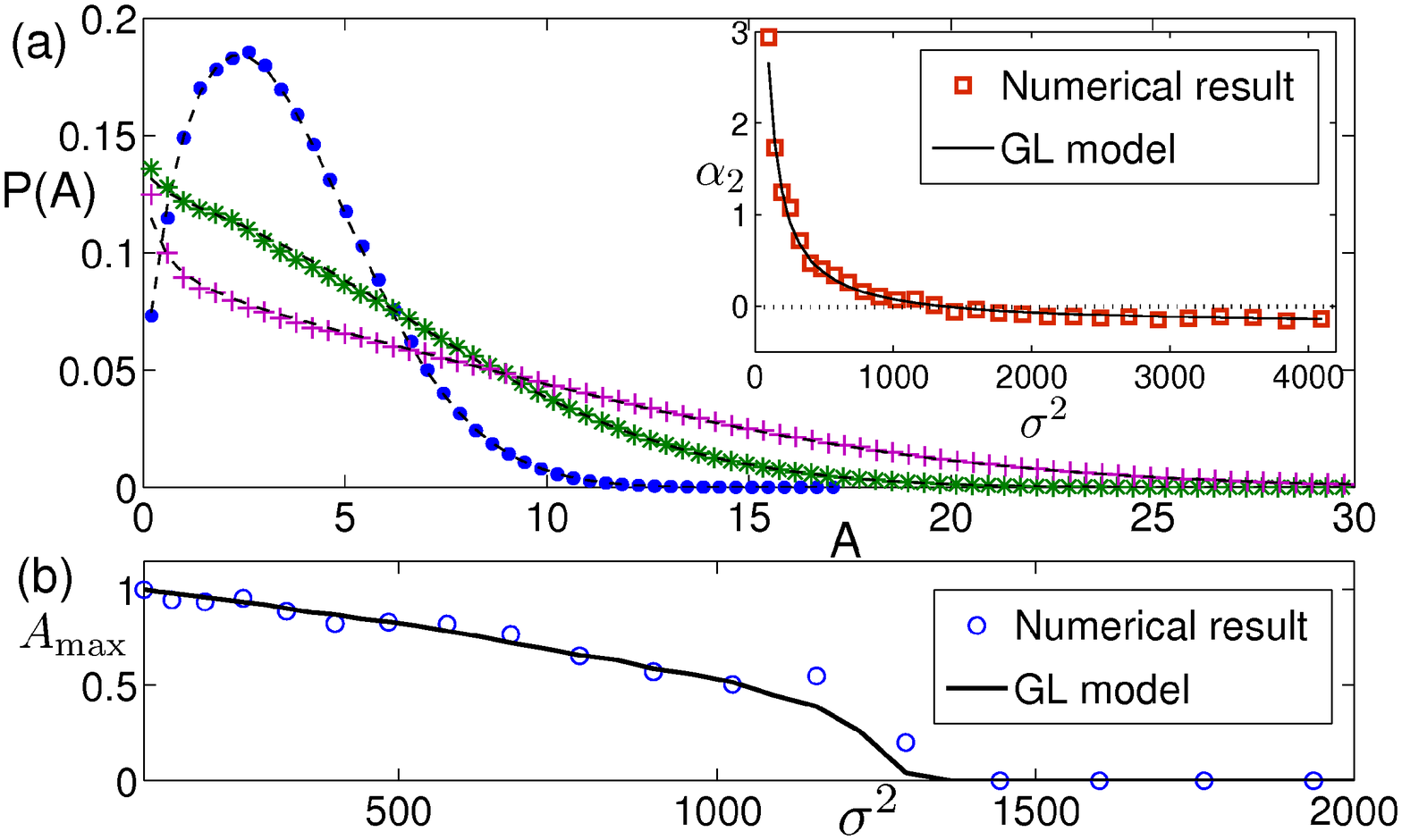}
\caption{(Color online) Numerical results for the stochastic KS
equation (\ref{Eq:KS}) integrated on a $[-\pi,\pi]$ domain with PBC.
(a) PDF of the first-mode amplitude for $\sigma=25$ ($\bullet$),
$45$ ($\ast$), and $65$ ($+$), with $\epsilon=0.025$. Dashed lines
correspond to a data fit using $P(A)=N A^{\alpha_2}\exp{(-\delta'
A^2)}$. The inset shows the value of the fitted exponent $\alpha_2$
as a function of $\sigma^2$ compared to the analytical solution
(solid line) given by  Eq.(\ref{Eq:alpha2}). (b) Maximum of the PDF
as a function of $\sigma^2$ normalized with the value corresponding at
$\sigma=10$. The solid line corresponds to the
numerical solution of Eqs.~(\ref{Eq:GL 2Da}) and
(\ref{Eq:GL 2Db}), by using Eq.~(\ref{Eq:coeff KS}). 
%All curves have been normalized with the corresponding minimum value at $\sigma=10$.
}\label{Fig:fig4}
\end{figure}%

When the system is solved by imposing PBC, the first state
transition occurs at $\sigma\simeq 36$ [cf.~Fig.~\ref{Fig:fig4}].
Interestingly, as we increase the noise strength, the second
critical transition is no longer observed, and  the power-law regime
of the amplitude PDF increases with an exponent that is asymptotically
decreasing up to the value around $-0.21$.
Finally, the same analysis for both DBC and PBC  has been performed
for the noisy Burgers equation (used, for example, as a prototype
for 1D turbulence, albeit without pressure
gradient)
%~\cite{Sreenivasan1999})
: $\partial_t
u=(\partial_x^2+1)u+\epsilon^2u+ u\partial_x u+\epsilon\sigma\xi$.
We obtain the  values $\sigma_\mathrm{I}\simeq 6.4$ and
$\sigma_\mathrm{II}\simeq 9.5$ for the DBC case, including  
on-off intermittency with the same exponent for the
waiting times' PDF (see Fig.~\ref{Fig:fig3}). In the PBC case, we
find $\sigma_\mathrm{I}\simeq 9.5$, and as  with the KS equation,
the transition to state~III is not observed either. This critical
phenomenon occurring in both models reflects an underlying
\emph{universal} behavior. Our aim now is to explain all these %the above
numerical results using multiscale analysis (singular perturbation
theory).
%~\cite{Blomker.etal_2007,PavlSt08}
%, appropriately modified
%so that it can be applied to SES such as the noisy KS and Burgers
%equations.

\emph{Multiscale theory.-} We first analyze the  noisy KS
equation~\cite{Applica}. In the limit of $\epsilon\ll 1$, the system
is close to the bifurcation point and Eq.~(\ref{Eq:KS}) has two
widely separated time scales, corresponding to the (stable) fast and
(unstable) slow modes. Considering then the behavior
of small solutions   at time scales of $O(\epsilon^{-2})$, we define
$u(x,t)=\epsilon v(x,\epsilon^2 t)$ %and use the scaling properties of white noise 
to transform Eq.~(\ref{Eq:KS}) to
\begin{equation}\label{Eq:KS scaled}
\partial_t v=-\epsilon^{-2}(\partial_x^2+\partial_x^4)v+
\partial_x^4 v-\epsilon^{-1}v\partial_x v+
\epsilon^{-1} \sigma e_2\hat{\xi}_2(t),
\end{equation}
where we have assumed highly degenerate noise. %As before, we shall
%consider separately both cases of boundary conditions.

For DBC, the solution can be expanded in the basis
$\left\{e_k(x)=c_k\sin{(kqx)}\right\}$, where $q=\pi/L$,  $c_k$'s
are normalization constants, and the single dominant mode
$\hat{u}_1(t)$ is real and belongs to the null space of
$\mathcal{L}=-\partial_x^2-\partial_x^4$. Also, we stipulate 
that $\xi(x,t)=\beta(t)\sin{(2qx)}$, with
$\beta(t)$ being  %uncorrelated 
white noise, $\langle \beta(t)
\beta(t')\rangle =\delta(t-t')$. To obtain the dominant mode 
amplitude equation we project the field $v$ in Eq.~(\ref{Eq:KS
scaled}) onto the null space of $\mathcal{L}$ to get
$w_1=\mathcal{P}_c v$,  where $\mathcal{P}_c$ is the corresponding
projector to the null space, and onto its orthogonal subspace
(stable modes) to get $w_\perp=(\mathcal{I}-\mathcal{P}_c) u$, where
$\mathcal{I}$ is the identity operator. Equation~\eqref{Eq:KS
scaled}, when written for the variables $w_1$ and $w_{\perp}$ is of
the form of a fast/slow system of stochastic differential equations
(SDEs) for which homogenization theory applies~\cite{PavlSt08}. By
analyzing the corresponding Fokker-Planck equation using singular
perturbation theory we obtain a closed equation for the distribution
function of $w_1$ from which we can read off a one-dimensional
stochastic differential equation that is valid in the limit $\eps
\to 0$. The resulting equation for $A(t)=\vert w_1\vert$ is given by
the GL equation with multiplicative Stratonovich noise:
\begin{equation}\label{Eq:GL}
\dot{A}=(1+\gamma_1\sigma^2) A-\gamma_2 A^3
+\gamma_3\sigma A\beta(t),
\end{equation}
where
\begin{equation}\label{Eq:coeff KS}
\gamma_1=-1/2688,\quad \gamma_2=1/48, \quad \gamma_3=1/24.
\end{equation}
Equation (\ref{Eq:GL}) has been the subject of several
studies (e.g.~see Ref.~\cite{Sagues.etal_2007} and references
therein), and the corresponding stationary PDF for the random
variable $A$ is found to be~\cite{Mackey.etal_1990}
$P(A)=N A^{\alpha_1}\exp{(-\delta A^2)}$, %\label{PDF}
with $N$ a normalization constant, and
\begin{equation}\label{Eq:paramters}
\alpha_1(\sigma)=2(1+\gamma_1\sigma^2)/(\gamma_3^2\sigma^2)-1,
\quad \delta(\sigma)=\gamma_2/(\gamma_3^2\sigma^2).
\end{equation}
As noted in
Ref.~\cite{Horsthemke_Lefever_1984}, depending on the location
of the maxima of the above PDF, there may exist different
states describing the amplitude $A$.
The interesting point 
is that all the numerical  states presented before can be achieved
by simply changing the  value of $\sigma$. First, we observe
that as long as $\alpha_1>0$ the maximum of $P(A)$ occurs at a
finite value, $A_\mathrm{max}>0$, and then $A$ is
characterized by finite fluctuations around a mean value (state I).
In contrast, for $-1<\alpha_1\leq 0$, the maximum is located at zero,
$A_\mathrm{max}=0$, and the amplitude fluctuates intermittently
between zero and a finite value (state II). These two states are
separated by the critical value:
\begin{equation}\label{Eq:sigma I}
\sigma_\mathrm{I}=(\gamma_3^2/2-\gamma_1)^{-1/2}.
%2[1+\gamma_1(\sigma_\mathrm{I})]=\gamma_3^2(\sigma_\mathrm{I})
\end{equation}
Note that for $\gamma_1>0$, this transition can only be observed
as long as $\gamma_3^2>2\gamma_1$, while it is always observed for
$\gamma_1<0$.  By using the values of Eq.~(\ref{Eq:coeff KS}) we therefore
obtain $\sigma_\mathrm{I}=28.4$ in excellent agreement with the numerical
observation shown at Fig.~\ref{Fig:fig2}$b$. In addition, the critical behavior can
be characterised as
$A_\mathrm{max}=
\vert\sigma_\mathrm{I}^2-\sigma^2\vert^{1/2}/(\sigma_\mathrm{I}\sqrt{\gamma_2})$
 for $\sigma \leq \sigma_\mathrm{I}$,
and $A_\mathrm{max}=0$ otherwise, so that $A_\mathrm{max}$ and
$\sigma^2$ are the order and control parameter, respectively,
describing the critical transition. By solving numerically
Eq.~(\ref{Eq:GL}) with the coefficients of Eq.~(\ref{Eq:coeff
KS}) for different $\sigma$, we find very good agreement between analytical 
and numerical results [cf.~Fig.~\ref{Fig:fig2}$b$].
If $\gamma_1<0$, a second transition occurs when $\alpha_1 \leq-1$.
The PDF cannot be normalized and it is given by a Dirac delta
function, $P(A)=\delta(A)$, describing a completely stabilized state
with $A=0$ (state III). The critical value $\sigma_\mathrm{II}$ for
this second transition is:
\begin{equation}\label{Eq:sigma II}
\sigma_\mathrm{II}=\sqrt{1/\vert\gamma_1\vert},
\end{equation}
yielding $\sigma_\mathrm{II}=51.8$, in excellent agreement with the
numerical results (cf. Fig.~\ref{Fig:fig2}$a$). To obtain
analytically the statistical properties of the waiting times $T$, we
assume that in a regime close to the critical point
($\sigma\lesssim\sigma_\mathrm{II}$) the initial value of $A$ is
below a small given threshold $c_\mathrm{th}$, and we ask for the
probability $P(T)$ that at time $T$ the amplitude reaches the
threshold for the first time. In this close-to-zero state we can
neglect the nonlinear term in
Eq.~(\ref{Eq:GL}), 
and we introduce the transformation $y=\log{A}$ obtaining
$\dot{y}=1+\gamma_1\sigma^2+\gamma_3\sigma\beta(t)$ with $y\in
(-\infty,\log{c_\mathrm{th}}]$. We  thus recognise an underlying
dynamics described by the well-known first-passage properties of the
random walk~\cite{FirstPassage_book}, giving rise in our case to the
long-time behavior  $P(T)\sim T^{-3/2}\exp{(-T/T_0)}$ with
$T_0=[2\gamma_3\sigma/(1+\gamma_1\sigma^2)]^2$, from which in the
critical point $\sigma=\sigma_\mathrm{II}$ we recover the
numerically observed pure power-law (cf.~Fig.~\ref{Fig:fig3}).
Clearly, the exponent $-3/2$ will be universally observed in any SES
whose dominant mode is described by Eq.~(\ref{Eq:GL}).

Consider now the case with PBC. The solution is then expanded in the
exponential Fourier basis
$\left\{e_k(x)=c_k\exp{(\mathrm{i}kqx)}\right\}$, for
$k=0,\pm1,\pm2,\ldots,$ and the single dominant mode has two
components: $\hat{u}_1(t)e_1(x)=y_1\sin(qx)+z_1\cos(qx)$. The noise
is now given as $\xi(x,t)=\beta_1(t)\sin{(2qx)} +
\beta_2(t)\cos{(2qx)}$, where $\beta_1(t)$ and $\beta_2(t)$ are
uncorrelated white random variables.
By applying our multiscale methodology %of the previous case,
we obtain:
%obtain that the dynamics for the two components of the
%first mode is given by:
\begin{align}
\dot{y}_1 & =  (1+2\gamma_1 \sigma^2) y_1-\gamma_2 y_1A^2+2\gamma_3\sigma A \beta_1, \label{Eq:GL 2Da} {} \\
\dot{z}_1 & =  (1+2\gamma_1 \sigma^2) z_1-\gamma_2 z_1A^2+2\gamma_3\sigma A \beta_2, \label{Eq:GL 2Db}
\end{align}
where  $A(t)=\sqrt{y_1^2+z_1^2}$, and
%the coefficients
$\gamma_1$, $\gamma_2$, and $\gamma_3$ are given by Eq.~(\ref{Eq:coeff KS}).
The stationary joint PDF for the two variables, $G(y_1,z_1)$, is %can be
obtained by computing the corresponding stationary two-dimensional
Fokker-Planck equation, yielding % This yields
%
%\begin{equation}\label{Eq:joint PDF}
$G(y_1, z_1) \propto(y_1^2+z_1^2)^{\alpha_1'/2}\exp{[-\delta'(y_1^2+z_1^2)]}$,
%\end{equation}
%
where %the parameters
$\alpha_1'$ and $\delta'$ are obtained from the
expressions in Eq.~(\ref{Eq:paramters}) by replacing $\gamma_1$ and
$\gamma_3$ with $2\gamma_1$ and $2\gamma_3$, respectively.
To study the behavior of $P(A)$, we move to a polar coordinate system
$(A,\theta)$
and impose the condition
$G(y_1,z_1)\mathrm{d}y_1\mathrm{d}z_1=P(A,\theta)\mathrm{d}A\mathrm{d}\theta$,
getting
$P(A)\propto A^{\alpha_2}\exp{(-\delta' A^2)}$,
with
\begin{equation}\label{Eq:alpha2}
\alpha_2(\sigma)=\alpha_1'+1=(1+2\gamma_1\sigma^2)/(2\gamma_3^2\sigma^2).
\end{equation}
We first note that state transitions  can only
occur iff $\gamma_1 <0$, with the critical values:
$\sigma_\mathrm{I}=\sqrt{1/2\vert\gamma_1\vert}$, and 
$\sigma_\mathrm{II}=[2(\vert\gamma_1\vert-\gamma_3^2)]^{-1/2}$.
Interestingly, the second transition can only occur as long as
$\gamma_3^2<\vert\gamma_1\vert$. Otherwise, %the completely stabilized 
state III is never observed, and the distribution tends
to $P(A)\sim A^{\alpha_\infty}$ as $\sigma\to\infty$, with
$\alpha_\infty=-\vert\gamma_1\vert/\gamma_3^2$.
By using Eq.~(\ref{Eq:coeff KS}), the first
transition occurs at $\sigma_\mathrm{I}=36.3$, while the second
transition cannot be observed with %giving the asymptotic value
$\alpha_\infty=-0.21$, in excellent agreement with the numerical
results (cf.~Fig.~\ref{Fig:fig4}).
Finally, when this formalism is applied to the Burgers equation, we
obtain the coefficients $\gamma_1=-1/88$, $\gamma_2=1/12$, and
$\gamma_3=1/6$, giving rise to %the critical points
$\sigma_\mathrm{I}=6.3$ and $\sigma_\mathrm{II}=9.4$ for DBC,
and $\sigma_\mathrm{I}=9.4$ for PBC,
in excellent agreement with the numerical results. As with the
KS equation, we have %the condition 
$\gamma_3^2>\vert \gamma_1\vert$, %does not hold, 
and the second transition is not observed for PBC either.

To conclude, we have presented clear evidence of critical
transitions in SES induced by pure additive noise. We have focused
on the KS equation
%randomly perturbed in the vicinity of a primary
%bifurcation point.
and by adding a stochastic forcing acting on the first stable mode,
we have provided a detailed and systematic investigation of the
transitions between different states. In particular, by using
multiscale analysis for SDEs, we have analytically described the
different critical-state transitions that are undergone by the
amplitude of the unstable mode, including on-off intermittency and
stabilized states, that we have also observed numerically in both
the KS and Burgers equations. Moreover, the critical exponents for
both SES are the same, and hence they belong to the same
universality class. This is in accordance with Yakhot's
conjecture~\cite{yakhot81}. We believe that our results will
motivate further analytical and numerical studies on the effect of
additive noise in general SES.

We thank Christian Ruyer-Quil for useful discussions. We acknowledge
financial support from EU-FP7 ITN Multiflow. DTP was partly
supported by NSF Grant DMS-0707339.


\begin{thebibliography}{99}
\bibitem{Applica}
The formalism is of much wider applicability and could be easily
extended to other prototypes.
\end{thebibliography}
\end{document}